\documentclass{appolb}
\usepackage{graphicx,amsmath}

\begin{document}
\title{Axial Anomaly and Light Cone Distributions%
\thanks{Presented by the first author at Light Cone 2012, Cracow, Polish Academy of Arts and Sciences, 8-13 July 2012}%
}
\author{Yaroslav Klopot$^{a}$\footnote{On leave from   Bogolyubov Institute for Theoretical Physics, Kiev,
Ukraine}, Armen Oganesian$^{a,b}$, Oleg Teryaev$^{a}$
\address{${}^{a}$Joint Institute for Nuclear Research, Dubna 141980, Russia \\
${}^{b}$Institute of Theoretical and Experimental Physics, Moscow 117218, Russia}
\\
}
\maketitle
\begin{abstract}
Axial anomaly leads to exact sum rules for transition form factors providing the important constraints to respective distribution amplitudes. This rigorous NPQCD approach is valid even if QCD factorization is broken. The status of possible small non-OPE corrections to continuum in comparison to BABAR and BELLE data is discussed.
\end{abstract}
  
\section{Anomaly sum rule and transition form factors}

The phenomenon of axial anomaly  \cite{Bell:1969ts} is widely known for its manifestation in  two-photon decays of pseudoscalar mesons. The dispersive approach to axial anomaly \cite{Horejsi:1994aj} turns out to be a useful tool for exploration of the processes, which involve virtual photons also, like the photon-meson transitions $\gamma \gamma^* \to \pi^0 (\eta, \eta')$ \cite{Klopot:2010ke,Klopot:2011qq,Klopot:2011ai,Melikhov:2012qp,Klopot:2012hd}.

The axial anomaly is associated with the VVA triangle graph amplitude, which involves  two vector currents with momenta $k,q$ and one axial current with momentum $p=k+q$:
 
\begin{equation} \label{VVA}
T_{\alpha \mu\nu}(k,q)=\int
d^4 x d^4 y e^{(ikx+iqy)} \langle 0|T\{ J_{\alpha 5}(0) J_\mu (x)
J_\nu(y) \}|0\rangle.
\end{equation}
This amplitude can be decomposed into the six tensor structures,

\begin{align}
\label{eq1} \nonumber T_{\alpha \mu \nu} (k,q)  & =  F_{1} \;
\varepsilon_{\alpha \mu \nu \rho} k^{\rho} + F_{2} \;
\varepsilon_{\alpha \mu \nu \rho} q^{\rho}
 + \;  F_{3} \; k_{\nu} \varepsilon_{\alpha \mu \rho \sigma}
k^{\rho} q^{\sigma} \\
  & + F_{4} \; q_{\nu} \varepsilon_{\alpha \mu
\rho \sigma} k^{\rho}
q^{\sigma} + \;  F_{5} \; k_{\mu} \varepsilon_{\alpha \nu
\rho \sigma} k^{\rho} q^{\sigma} + F_{6} \; q_{\mu}
\varepsilon_{\alpha \nu \rho \sigma} k^{\rho} q^{\sigma},
\end{align}
where  $F_{j} = F_{j}(p^{2},k^{2},q^{2}; m^{2})$, $j = 1, \dots ,6$ are the scalar factors, constrained by current conservation and Bose symmetry.  In what follows, we consider the case with one virtual photon ($-q^2=Q^2>0$) and one real photon ($k^2=0$).

The axial anomaly, considered in the dispersive approach, leads to an anomaly sum rule (ASR)  \cite{Horejsi:1994aj}, 
\begin{equation}\label{asr}
\int_{4m^{2}}^{\infty} A_{3}^{(a)}(s,Q^{2}; m^{2}) ds =
\frac{1}{2\pi}N_c C^{(a)}, \; a=3,8,
\end{equation}
where $A_{3} = \frac{1}{2}Im_{p^2} (F_3-F_6)$, $N_c=3$ is a number of colors, $m$ is a quark mass and $ C^{(a)}$ are the charge factors of components of the axial currents $J_{\alpha 5}^{(a)}$.  For the isovector ($a=3$)  and octet ($a=8$) components of axial current

\begin{align} 
J^{(3)}_{\mu 5}&=\frac{1}{\sqrt{2}}(\bar{u} \gamma_{\mu} \gamma_5 u - \bar{d}
\gamma_{\mu} \gamma_5 d ),\;\; C^{(3)}=\frac{1}{3\sqrt{2}},  \nonumber\\
J^{(8)}_{\mu 5}&=\frac{1}{\sqrt{6}}(\bar{u} \gamma_{\mu} \gamma_5 u + \bar{d}
\gamma_{\mu} \gamma_5 d - 2\bar{s} \gamma_{\mu} \gamma_5 s),\;\;
 C^{(8)}=\frac{1}{3\sqrt{6}},
\end{align}
the ASR (\ref{asr}) has an important property -- both perturbative and nonperturbative corrections to the integral are absent because of the Adler-Bardeen theorem and the 't Hooft's principle.


In the case of \emph{isovector channel}, saturating the  l.h.s. of the three-point correlation function (\ref{VVA}) with the resonances, singling out
the first contribution, given by the pion, and collecting all the other states into the continuum contribution $I_{cont}^{(3)}(s_3,Q^2)$,  we get the ASR in a form (in what follows we take $m=0$):  

\begin{equation} \label{qhd3}
\pi f_{\pi}F_{\pi\gamma}(Q^2)+ I_{cont}^{(3)} (s_3,Q^2) =\frac{1}{2\pi}N_c C^{(3)}, I_{cont}^{(3)}\equiv \int_{s_3}^{\infty} A_{3}^{(3)}(s,Q^{2};m^{2}) ds, 
\end{equation}
where $s_3$ is a continuum threshold, and the general definitions of the decay constants $f_M^{a}$ ($f_{\pi}^{(3)}\equiv f_\pi=130.7$ MeV) and  the transition form factors (TFFs) of the reactions $\gamma\gamma^* \to M$   are
\begin{equation}\label{def_f}
\langle 0|J^{(a)}_{\alpha 5}(0) |M(p)\rangle=
i p_\alpha f^a_M, \int d^{4}x e^{ikx} \langle M(p)|T\{J_\mu (x) J_\nu(0)
\}|0\rangle = \epsilon_{\mu\nu\rho\sigma}k^\rho q^\sigma
F_{M\gamma}. 
\end{equation}

If we employ the one-loop expression for the spectral density \cite{Horejsi:1994aj}
\begin{equation} \label{a3}
A_{3}^{(3)}(s,Q^{2})=\frac{N_c C^{(3)}}{2\pi}\frac{Q^2}{(Q^2+s)^2},
\end{equation}
from the Eq. (\ref{qhd3}) we get \cite{Klopot:2010ke}
\begin{align} \label{f3m}
F_{\pi\gamma}(Q^2)=\frac{1}{2\sqrt{2}\pi^2f_{\pi}}\frac{s_3}{s_3+Q^2}.
\end{align}

In the QCD factorization  approach the expression of the TFF is given in terms of the convolution of a hard scattering kernel and a soft pion distribution amplitude (DA) $\phi(x)$ (see e.g. \cite{Bakulev:2012nh} and references therein). In particular,  at $Q^2\to \infty$, where the pion DA evolves to its asymptotic form $\phi(x)^{as}=6x(1-x)$ and  the pion TFF acquires its asymptotic value \cite{Lepage:1980fj} $Q^2F_{\pi\gamma}^{as}(Q^2)=\sqrt{2}f_\pi$, the continuum threshold  $s_3$ can be determined from (\ref{f3m}), $s_3=4\pi^2f_\pi^2=0.67$ GeV$^2$ and then (\ref{f3m}) reproduces a well-known Brodsky-Lepage interpolation formula \cite{Brodsky:1981rp}.

\begin{figure}[htb]
\centerline{%
\includegraphics[width=9.0cm]{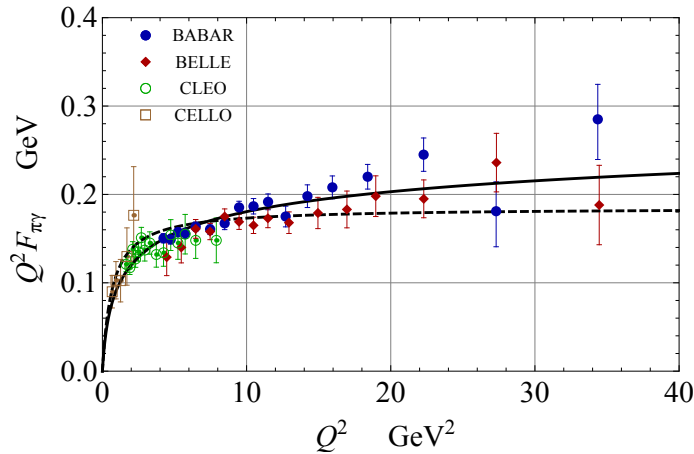}}
\caption{Pion transition form factor: Eqs. (\ref{f3m}) (dashed line) and (\ref{corr3F}) (solid line) compared with experimental data}
\label{fig:1}
\end{figure}

When compared to the experimental data on pion TFF, the equation (\ref{f3m}) gives a fairly good description of the data of CELLO \cite{Behrend:1990sr}, CLEO \cite{Gronberg:1997fj} and BELLE \cite{Uehara:2012ag} collaborations, while the data of BABAR collaboration \cite{Aubert:2009mc} are described much worse\footnote{The similar result is obtained also in the  LCSR approach \cite{Bakulev:2012nh}, where it was shown that the BABAR data cannot be satisfactory described with only two Gegenbauer coefficients.} (see dashed line in Fig. \ref{fig:1}). The BABAR data indicate a log-like growth, and in order to describe them well, one needs to consider a possibility of the correction. As we mentioned above, the  integral in the ASR does not have any corrections, but the spectral density $A_{3}^{(3)}(s,Q^{2})$ can acquire corrections, and therefore the continuum and the pion contributions can  have  corrections as well. The exactness of the ASR results in an interesting interplay between corrections to the continuum and pion: they should cancel each other to preserve the ASR, $\delta I_{cont}^{(3)}=-\delta I_\pi$. The form of the correction is not yet known (the  origins of such a correction should be essentially nonperturbative, see discussion in \cite{Klopot:2012hd}). Nevertheless, we can propose the form of the correction, relying on  general properties of ASR: it should vanish at $s_3\to \infty$ (the continuum contribution vanishes), at  $s_3\to 0$ (the full integral has no corrections), at  $Q^2\to \infty$ (the perturbative theory works at large $Q^2$)  and at $Q^2\to 0$ (anomaly perfectly describes pion decay width).  Supposing the correction contains rational functions and logarithms of $Q^2$, the simplest form of the correction satisfying those limits results  \cite{Klopot:2012hd} in 

\begin{align} \label{corr3F}
F_{\pi\gamma}(Q^2) = \frac{1}{\pi f_\pi} (I_\pi + \delta I_\pi)  = \frac{1}{2\sqrt{2}\pi^2f_{\pi}}\frac{s_3}{s_3+Q^2}\Bigl [1+\frac{\lambda Q^2}{s_3+Q^2}&(\ln{\frac{Q^2}{s_3}}+\sigma)\Bigr ],
\end{align}  
where $\lambda$ and $\sigma$ are dimensionless parameters. This kind of correction cannot  appear in (a local) OPE and should be attributed, possibly, to instantons or short strings. Note also, that this correction implies that the pion distribution amplitude $\phi(x)$ does not vanish at $x=0,1$ and violates the factorization (see also \cite{Radyushkin:2009zg,Polyakov:2009je}).

The fit of the TFF (\ref{corr3F}) to the combined CELLO+CLEO+BABAR data gives $\lambda=0.14,\;\sigma=-2.36$, $\chi^2/d.o.f. =0.94 \;\; d.o.f.=35$. The plot of $Q^2F_{\pi\gamma}$  for these parameters is shown in Fig. \ref{fig:1} as a solid line. 
The TFF  (\ref{corr3F}) with these parameters $\lambda,\sigma$  describes well also the combined CELLO+CLEO+BELLE data with $\chi^2/d.o.f. =0.84 \; (d.o.f.=35)$. On the other hand, the TFF without correction (\ref{f3m}) (dashed line in Fig. \ref{fig:1})) gives $\chi^2/d.o.f. =2.29$ and  $\chi^2/d.o.f. =1.01$ for CELLO+CLEO+BABAR and CELLO+CLEO+BELLE data sets respectively.
We can conclude, that although the BABAR data favour the log-like correction, the newly released BELLE data neither confirm, nor exclude the possibility of this correction.

It is interesting to consider in the same way the ASR in the \emph{octet channel}. Here we should take into account the first two contributions, which are given by $\eta$ and $\eta'$ mesons. Then the ASR in the octet channel \cite{Klopot:2011qq} is  (cf. also \cite{Feldmann:1998yc}):

\begin{equation} \label{asr8}
f_{\eta}^8 F_{\eta\gamma}(Q^2) +f_{\eta'}^8F_{\eta'\gamma}(Q^2)=  \frac{1}{2\sqrt{6}\pi^2}\frac{s_8}{s_8+Q^2},
\end{equation}
where $s_8$ is a continuum threshold, which can be determined from the large-$Q^2$ limit of (\ref{asr8}) and the pQCD  predicted expression for the $\eta, \eta'$ TFFs:  
\begin{equation} \label{asr8inf}
s_8=4\pi^2((f_\eta^8)^2+(f_{\eta'}^8)^2+ 2\sqrt{2} [ f_\eta^8 f_{\eta}^0+ f_{\eta'}^8 f_{\eta'}^0]).
\end{equation}

Naturally, if the log-like correction is present in the isovector channel, it should reveal itself in the octet channel too. The similar correction in the octet channel leads to the ASR with the correction term \cite{Klopot:2011ai,Klopot:2012hd}:

\begin{align} \label{asr8-corr}
f_{\eta}^8 F_{\eta\gamma}(Q^2) +f_{\eta'}^8F_{\eta'\gamma}(Q^2)=  \frac{1}{2\sqrt{6}\pi^2}\frac{s_8}{s_8+Q^2}\Bigl [1+\frac{\lambda Q^2}{s_8+Q^2}(\ln{\frac{Q^2}{s_8}}+\sigma)\Bigr ].
\end{align}

\begin{figure}[htb]
\centerline{%
\includegraphics[width=9.0cm]{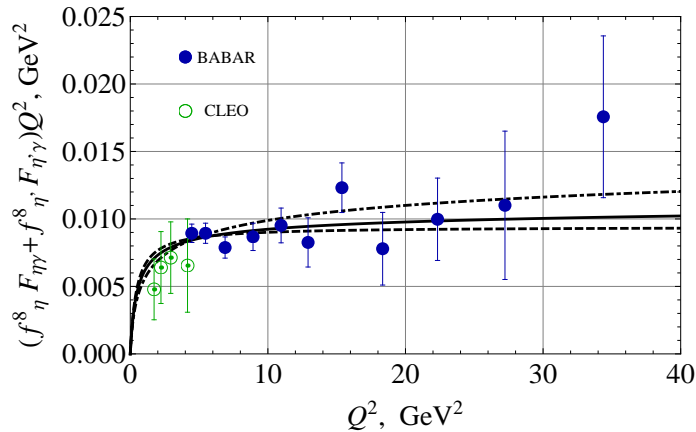}}
\caption{The ASR in the octet channel for different values of fitting parameters compared with the experimental data, see description in the text.}
\label{fig:2}
\end{figure}

The Eqs. (\ref{asr8}), (\ref{asr8inf}) and (\ref{asr8-corr}) contain the decay constants $f_M^a$, which are usually analyzed basing on different mixing schemes or in a scheme independent way (see, e.g., \cite{Klopot:2012hd,Thomas:2007uy} and references therein). For the purposes of  numerical analysis, we employ the decay constants, obtained in a scheme-independent way in Ref. \cite{Klopot:2012hd}: $f_\eta^8=1.11f_\pi, f_{\eta'}^8=-0.42f_\pi,f_\eta^0=0.16f_\pi, f_{\eta'}^8=1.04f_\pi$.  Then, the fit of the Eq. (\ref{asr8-corr}) to the experimental data of BABAR collaboration \cite{BABAR:2011ad} gives $\lambda=0.05, \sigma=-2.58$ with $\chi^2/d.o.f.=0.81$ (see the solid line in Fig. \ref{fig:2}), while Eq. (\ref{asr8}) gives $\chi^2/d.o.f.=0.85$ (dashed line). At the same time, if the parameters are taken the same as for the pion case $\lambda=0.14, \sigma=-2.36$, we get $\chi^2/d.o.f.=1.02$ (dot-dashed line). We see that the current precision of the  experimental data on $\eta, \eta'$ TFFs can accommodate the log-like correction in the octet channel, although does not require it.    

\section{Conclusions}

The current experimental data for the pion transition form factor in the range of $Q^2=10-35 GeV^2$ available from BABAR \cite{BABAR:2011ad}  and BELLE \cite{Uehara:2012ag} collaborations manifest different tendencies.

The BABAR data \cite{BABAR:2011ad} show an excess over the asymptotic value of the transition form factor, requiring a log-like correction, and therefore violating the QCD factorization and favouring the flat-like (not vanishing at the edges) pion distribution amplitude.
The more recent BELLE data \cite{Uehara:2012ag} do not manifest that striking behaviour and are more or less consistent with the Brodsky-Lepage interpolation formula.
The  analysis for the octet channel of the ASR, based on the BABAR data on $\eta$ and $\eta'$ TFFs \cite{BABAR:2011ad}, shows the possibility to accommodate such correction, but does not require it.

Y.K. would like to thank the organizers and, in particular,  Wojciech Broniowski  for hospitality and a nice atmosphere at the meeting. The partial support from the Russian Foundation for Basic Research (grants 12-02-00613a, 12-02-00284a) and the Bogoliubov-Infeld program is acknowledged.


\begin{thebibliography}{55}



\bibitem{Bell:1969ts}
  J.~S.~Bell, R.~Jackiw,
  Nuovo Cim.\  {\bf A60}, 47-61 (1969).

  S.~L.~Adler,
  Phys.\ Rev.\  {\bf 177}, 2426-2438 (1969).



\bibitem{Horejsi:1994aj}
  J.~Horejsi, O.~Teryaev,
  Z.\ Phys.\  {\bf C65}, 691-696 (1995).



\bibitem{Klopot:2010ke} 
  Y.~N.~Klopot, A.~G.~Oganesian and O.~V.~Teryaev,
  Phys.\ Lett.\ B {\bf 695}, 130 (2011);

\bibitem{Klopot:2011qq} 
  Y.~N.~Klopot, A.~G.~Oganesian and O.~V.~Teryaev,
  Phys.\ Rev.\ D {\bf 84}, 051901 (2011);

\bibitem{Klopot:2011ai} 
  Y.~Klopot, A.~Oganesian and O.~Teryaev,
  JETP Lett.\  {\bf 94}, 729 (2011).

\bibitem{Melikhov:2012qp} 
  D.~Melikhov and B.~Stech,
  Phys.\ Lett.\ B {\bf 718}, 488 (2012)
  

\bibitem{Klopot:2012hd} 
  Y.~Klopot, A.~Oganesian and O.~Teryaev,
  arXiv:1211.0874 [hep-ph].

  \bibitem{Bakulev:2012nh} 
    A.~P.~Bakulev, S.~V.~Mikhailov, A.~V.~Pimikov and N.~G.~Stefanis,
    Phys.\ Rev.\ D {\bf 86}, 031501 (2012)
 
  
\bibitem{Lepage:1980fj} 
  G.~P.~Lepage and S.~J.~Brodsky,
  Phys.\ Rev.\  {\bf D22}, 2157 (1980).
 
 
\bibitem{Brodsky:1981rp}
  S.~J.~Brodsky, G.~P.~Lepage,
  Phys.\ Rev.\  {\bf D24}, 1808 (1981).

\bibitem{Behrend:1990sr} 
  H.~J.~Behrend {\it et al.}  [CELLO Collaboration],
  Z.\ Phys.\ C {\bf 49}, 401 (1991).

\bibitem{Gronberg:1997fj} 
  J.~Gronberg {\it et al.}  [CLEO Collaboration],
  Phys.\ Rev.\ D {\bf 57}, 33 (1998)
  

\bibitem{Uehara:2012ag}
  S.~Uehara {\it et al.}  [Belle Collaboration],
  arXiv:1205.3249 [hep-ex].

\bibitem{Aubert:2009mc} 
  B.~Aubert {\it et al.}  [BABAR Collaboration],
  Phys.\ Rev.\ D {\bf 80}, 052002 (2009)
  

\bibitem{Radyushkin:2009zg} 
  A.~V.~Radyushkin,
  Phys.\ Rev.\ D {\bf 80}, 094009 (2009)
  
  \bibitem{Polyakov:2009je} 
    M.~V.~Polyakov,
    JETP Lett.\  {\bf 90}, 228 (2009)


\bibitem{Feldmann:1998yc} 
  T.~Feldmann and P.~Kroll,
  Phys.\ Rev.\ D {\bf 58}, 057501 (1998)
  [hep-ph/9805294].


\bibitem{Thomas:2007uy} 
  C.~E.~Thomas,
  JHEP {\bf 0710}, 026 (2007)
  

\bibitem{BABAR:2011ad} 
  P.~del Amo Sanchez {\it et al.}  [BABAR Collaboration],
  Phys.\ Rev.\ D {\bf 84}, 052001 (2011)
  

\end{thebibliography}
\end{document}